\documentclass[journal,letterpaper,12pt,oneside,onecolumn,draftclsnofoot]{IEEEtran}
\usepackage{amsfonts}

\usepackage[dvips]{graphicx}
\usepackage{amsmath}
\usepackage{amssymb}
\usepackage{multirow}
\usepackage{longtable}
\usepackage{subfigure}
\usepackage{float}
\usepackage{afterpage}
\usepackage{multirow}
\usepackage[nospace]{cite}
\usepackage{subeqnarray}
\usepackage{algorithm}
\usepackage{algpseudocode}
\usepackage{mathrsfs}
\usepackage{xcolor}
\long\def\symbolfootnote[#1]#2{\begingroup
\def\thefootnote{\fnsymbol{footnote}}
\footnote[#1]{#2}\endgroup}
\IEEEoverridecommandlockouts

\begin{document}
%
% paper title
\title{\Large \bf Simulation Results of User Behavior-Aware Scheduling Based on Time-Frequency Resource Conversion} %\Large \bf
%, the National Natural Science Foundation
%of China (No. 60802011), and Huawei corporation.}
%\vspace{-0.3cm}
\author{\small \authorblockN{Hangguan Shan$^1$, Yani Zhang$^1$, Weihua Zhuang$^2$, Aiping Huang$^1$, and Zhaoyang Zhang$^1$}% \small
\\ \small \authorblockA{$^1$College of Information Science and Electronic Engineering, %\footnotesize
      Zhejiang University, Hangzhou, China
      }
\\ \authorblockA{$^2$Department of Electrical and Computer Engineering,
      University of Waterloo, Waterloo, Canada
      }

}
\maketitle%\symbolfootnote[0]{$^\dagger$This work was supported by National Natural Science Foundation of China (No. 61201228), Specialized Research Fund for the Doctoral Program of Higher Education (No. 20120101120077), and Zhejiang Provincial Natural Science Foundation of China (No. LY12F01021).}

\vspace{-0.30cm}
\begin{abstract}

Integrating time-frequency resource conversion (TFRC), a new network resource allocation strategy, with call admission control can not only increase the cell capacity but also reduce network congestion effectively. However, the optimal setting of TFRC-oriented call admission control suffers from the curse of dimensionality, due to Markov chain-based optimization in a high-dimensional space.
To address the scalability issue of TFRC, in \cite{Shan_TWC_submitted} we extend the study of TFRC into the area of scheduling.
Specifically, we study downlink scheduling based on TFRC for an LTE-type cellular network, to maximize service delivery.
The service scheduling of interest is formulated as a joint request, channel and slot allocation problem which is NP-hard.
An offline deflation and sequential fixing based algorithm (named DSFRB) with only polynomial-time complexity is proposed to solve the problem.
For practical online implementation, two TFRC-enabled low-complexity algorithms, modified Smith ratio algorithm (named MSR) and modified exponential capacity algorithm (named MEC), are proposed as well.
In this report, we present detailed numerical results of the proposed offline and online algorithms, which not only show the effectiveness of the proposed algorithms but also\emph{} corroborate the advantages of the proposed TFRC-based schedule techniques in terms of quality-of-service (QoS) provisioning for each user and revenue improvement for a service operator.

\textbf{Keywords:} Time-frequency resource conversion, context-aware resource allocation, scheduling.
\end{abstract}

%\vspace{0.3cm}
%%%%%%%%%%%%%%%%%%%%%%%%%%%%%%%%%%%%
% Introduction
%%%%%%%%%%%%%%%%%%%%%%%%%%%%%%%%%%%%
%\section{Introduction}
%\label{sec:introduction}
%
%Mobile data traffic has been grown in a explosive way. Explosive traffic
%
\section{Scheduling Performance}
%In this report, we evaluate the performance of the proposed TFRC-enabled scheduling techniques.

To evaluate the proposed TFRC-enabled scheduling techniques,  extensive simulations are performed for both offline and online scheduling.
For offline scheduling, we focus on the performance improvement of TFRC-enabled schedule and the benefits of the new penalty function proposed in \cite{Shan_TWC_submitted}, while for online scheduling we measure the performance gap between the proposed offline and multiple online algorithms.

Specifically, for offline scheduling, two benchmark algorithms are compared with the proposed DSFRB algorithm: the algorithm proposed in \cite{1zhang} and integrated with the new penalty function (denoted as ``DSF-NP") and the algorithm with the traditional penalty function (denoted as ``SF-OP")\footnote{Scheduling with the traditional penalty function is to solve a mixed integer linear OP (see Proposition 2 of \cite{Shan_TWC_submitted}). So, we can solve the OP directly with the SF algorithm proposed in \cite{Shan_TWC_submitted} effectively.}.
For online scheduling, two other benchmark algorithms are compared with the proposed MSR and MEC algorithms: the algorithm with earliest-deadline-first policy (denoted as ``EDF") and the algorithm proposed in \cite{Wu_WCNC13} (denoted as ``L-MaxWeight"). L-MaxWeight is tailored for scheduling flows with deadlines and is shown superior performance in underloaded identical-deadline systems.
For either  scenario, the impacts of different parameters, including traffic load, request size, mean request lifetime, and channel condition, on scheduling performance are studied.

In the simulation, we assume that there are 5 users being severed by a cell with 32 subchannels, each channel with a bandwidth of 15 KHz.
The channel fading is modeled by the Rayleigh distribution.
Each user initiates new requests according to a Poisson process with rate ${{\lambda }_{u}}$.
The data size $Q_k$ and lifetime ${\Pi }_{k}$ of the $k^{th}$ request are uniformly distributed within $\left[ {{Q}_{\min }},{{Q}_{\max }} \right]$ and  exponentially distributed with mean $\mu$, respectively.
The reward of the request's unit data ($A_k$) is assumed to be uniformly distributed within $[1, 10]$.
%The slot duration is set to be 100 ms.
The context information on user behavior is fed back from user equipment (UE) to the base station (BS) once every slot of 100 ms.
For all simulation results, we perform the simulation for 200 runs, average the results, and obtain the 95\% confidence intervals.

\subsection{Offline Scheduling}

\begin{figure*}[tp]%Ë«À¼Í¼Æ¬¼Ó*
\subfigcapskip = -7.5pt
\centering
\subfigure[Total reward vs. traffic load.]
{
\includegraphics[width=0.46\linewidth]{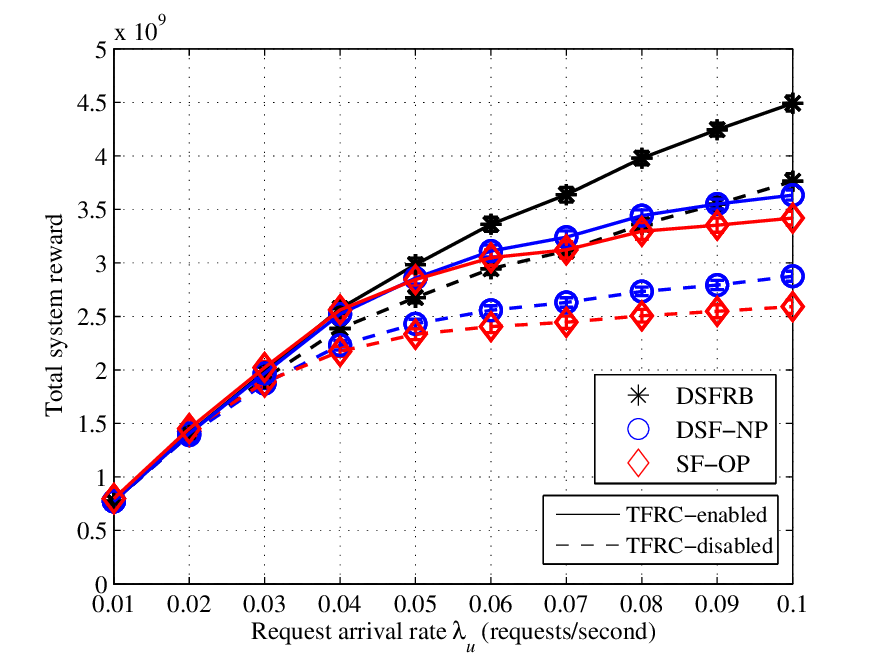}
\label{fig:lambda offline:reward}
}%
%\hspace{0.25in}
\subfigure[Complete ratio vs. traffic load.]
{
\includegraphics[width=0.46\linewidth]{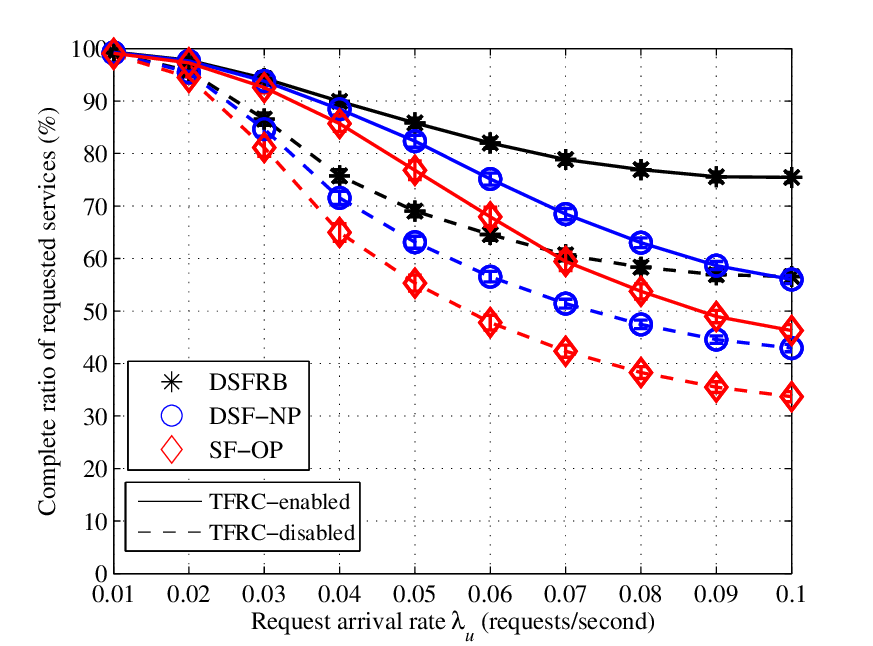}
\label{fig:lambda offline:ratio}
}
\vspace{-0.3cm}
\caption{Impact of traffic load on total system reward and complete ratio with offline scheduling.}
\vspace{-0.5cm}
\label{fig:lambda offline}
\end{figure*}

Fig. \ref{fig:lambda offline} shows the impact of traffic load on both total system reward (operator side) and complete ratio of all requests (user side),
with ${{\lambda }_{u}}\in \left[ 0.01,0.1 \right]$ requests/s, ${{Q}_{\min }}=15$ Mbits, ${{Q}_{\max }}=20$ Mbits,  $\mu =30$ s, and mean signal-to-noise ratio (SNR) for each channel equal to 10 dB.
Unless otherwise specified, simulation for effects of other parameters are all based on the same setting, and the changed parameters are listed.
In Fig. \ref{fig:lambda offline}, both results for the TFRC-enabled and TFRC-disabled DSFRB, DSF-NP, and SF-OP algorithms are presented.
It is observed that the algorithms enabling TFRC outperform their counterpart disabling TFRC, in terms of the system reward and complete ratio, generating a potential win-win situation for both the operator and the user.
The performance improvement results from the fact that the context information (which connection(s) is currently focused on by the user) utilized in the TFRC-enabled schedule offers more freedom to find a better scheduling solution.
Further, the performance gap, especially the one for DSFRB, increases with traffic load, illustrating the potential advantage of TFRC-enabled schedule in addressing a heavy-load network scenario.
In Fig. \ref{fig:lambda offline}, a performance gap between DSFRB and DSF-NP exists, mainly due to the newly proposed deflation strategy.
Both DSFRB and DSF-NP outperform SF-OP, illustrating an advantage of the new penalty function over the traditional one in  increasing delivered request number.

%DSFRB and DSF algorithm both perform better than SF algorithm in terms of the system reward and complete ratio for requested services, mainly due to the two different penalty methods. As mentioned in proposition 2, the old penalty method is equivalent to maximize the network's per-unit reward weighted throughput. However, the new penalty only charge reward for completed requests, therefore an offline scheduler will only schedule the requests which can be completed before deadline. Besides, the performance gap between DSFRB and DSF algorithm illustrate the advantage of the DSFRB's modification on DSF algorithm, especially the steps 4 to 6, which guarantee the reward improvement in each iteration. Further, we can see that all these algorithms enabling TFRC outperform their counterpart disabling TRFC, in both system reward and complete ratio, generating a win-win situation for both the operator and the user.

%\vspace{-0.15cm}

\begin{figure*}[tp]%Ë«À¼Í¼Æ¬¼Ó*
\subfigcapskip = -7.5pt
\centering
\subfigure[Total reward vs. average request size.]
{
\includegraphics[width=0.46\linewidth]{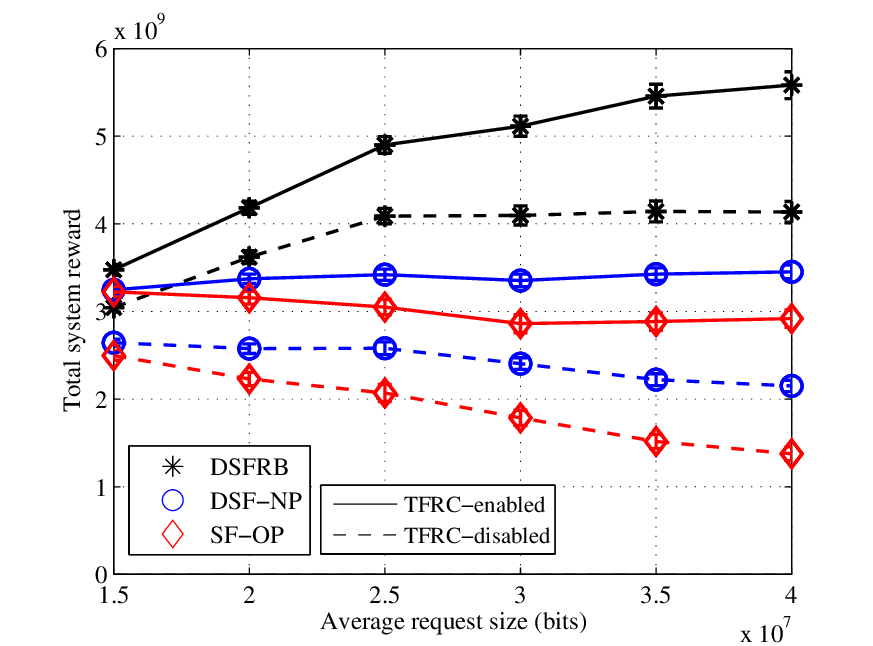}
\label{fig:data offline:reward}
}%
%\hspace{0.25in}
\subfigure[Complete ratio vs. average request size.]
{
\includegraphics[width=0.46\linewidth]{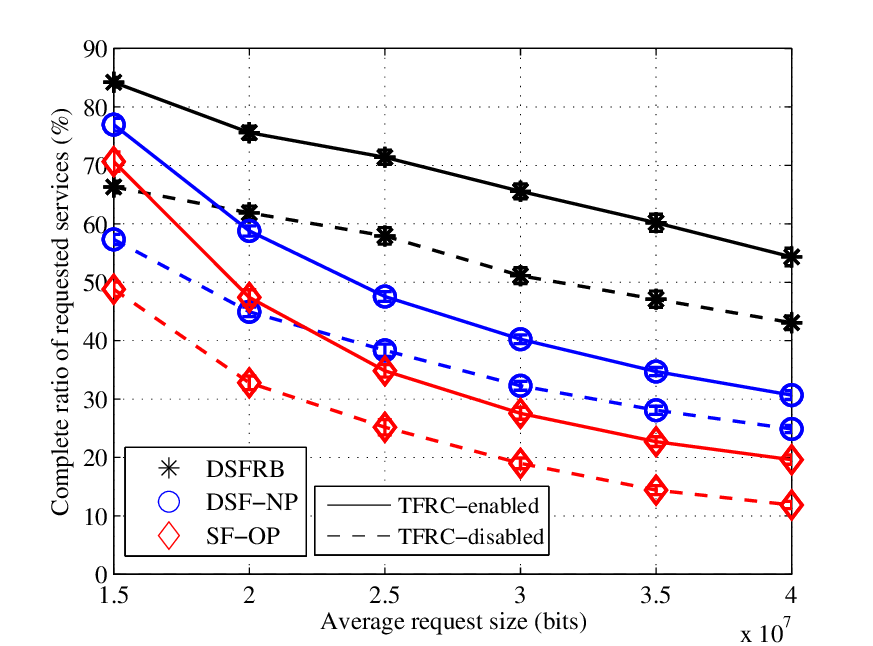}
\label{fig:data offline:ratio}
}
\vspace{-0.3cm}
\caption{Impact of average request size on total system reward and complete ratio with offline scheduling.}
\vspace{-0.65cm}
\label{fig:data offline}
\end{figure*}

Fig. \ref{fig:data offline} shows how the total system reward and the complete ratio change with the average request size, with ${{\lambda }_{u}}=0.075$ requests/s, ${{Q}_{\min }}=10$ Mbits,  and ${{Q}_{\max}}$ changing with the average request size from 15 Mbits to 40 Mbits. %, and other parameters set the same as for Fig. \ref{fig:lambda offline}.
We can see that the complete ratio of request services with any of the three algorithms decreases with an increase of the average request size, as the traffic load increases with the average request size.
The TFRC strategy makes each algorithm avoid suffering too much from the increased traffic load, as observed in Fig. \ref{fig:lambda offline}.
Further, it is observed in Fig. \ref{fig:data offline:reward} that, without applying TFRC, the total system reward for DSF-NP or SF-OP decreases with the average request size; however, the performance of algorithms enabling TFRC remains almost unchanged. %, illustrating the benefit of TFRC.
On the other hand, the total system reward for DSFRB without TFRC increases slightly for the average request sizes; yet, a much larger increasing rate is observed for the algorithm enabling TFRC. %showing not only the benefit of TFRC but also the superiority of DSFRB.

%we can see that the total system reward of SF-OP decreases with the increase of the traffic load. However, the performance of DSFRB, especially with TFRC strategy, increases slightly, showing the advantage of our newly proposed penalty method and DSFRB algorithm.

%\vspace{-0.1cm}
\begin{figure*}[tp]%Ë«À¼Í¼Æ¬¼Ó*
\subfigcapskip = -7.5pt
\centering
\subfigure[Total reward vs. mean request lifetime.]
{
\includegraphics[width=0.46\linewidth]{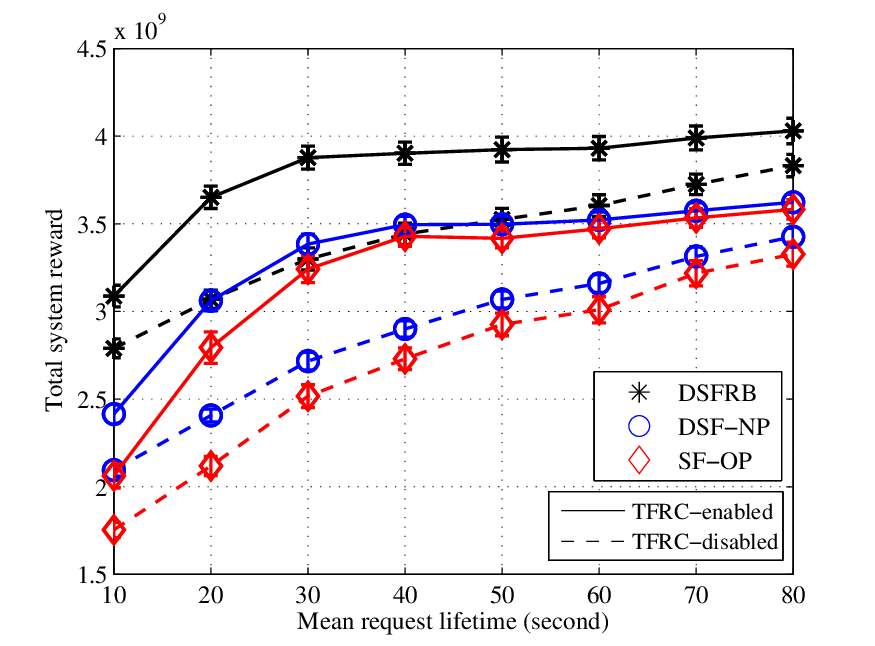}
\label{fig:lifetime offline:reward}
}%
%\hspace{0.25in}
\subfigure[Complete ratio vs. mean request lifetime.]
{
\includegraphics[width=0.46\linewidth]{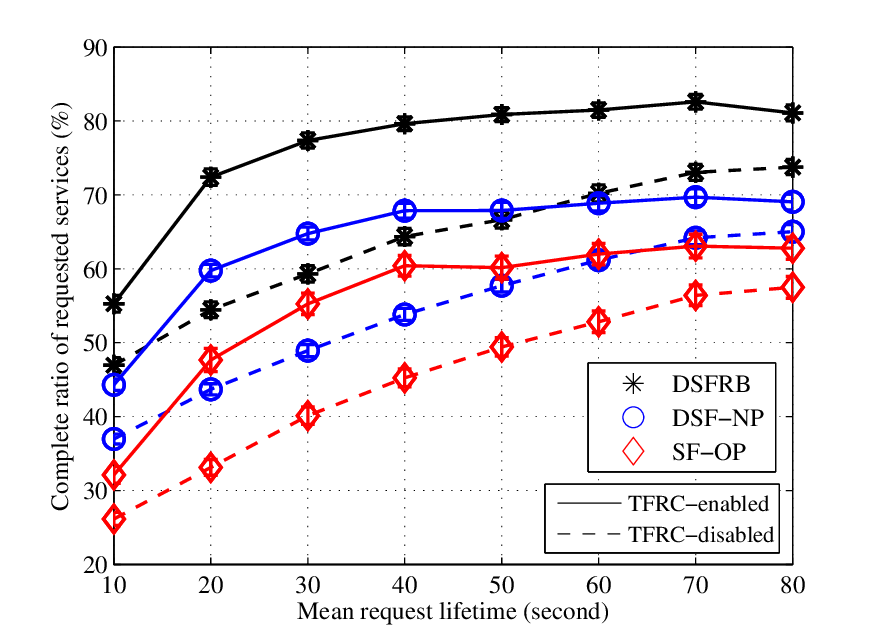}
\label{fig:lifetime offline:ratio}
}
\vspace{-0.3cm}
\caption{Impact of mean request lifetime on total system reward and complete ratio with offline scheduling.}
\vspace{-0.35cm}
\label{fig:lifetime offline}
\end{figure*}

\vspace{-0.1cm}
\begin{figure*}[tp]%Ë«À¼Í¼Æ¬¼Ó*
\subfigcapskip = -7.5pt
\centering
\subfigure[Total reward vs. mean SNR.]
{
\includegraphics[width=0.46\linewidth]{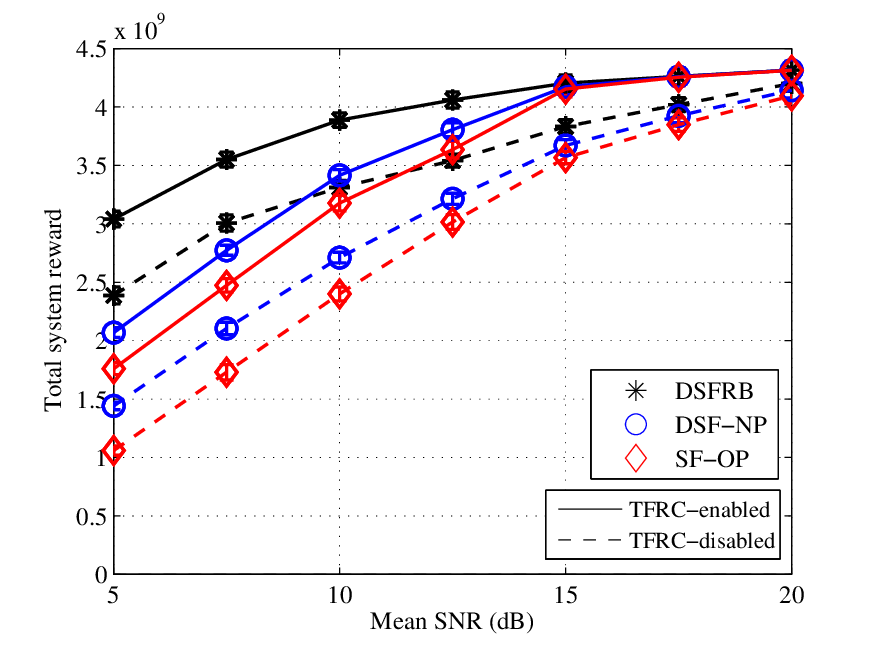}
\label{fig:snr offline:reward}
}%
%\hspace{0.25in}
\subfigure[Complete ratio vs. mean SNR.]
{
\includegraphics[width=0.46\linewidth]{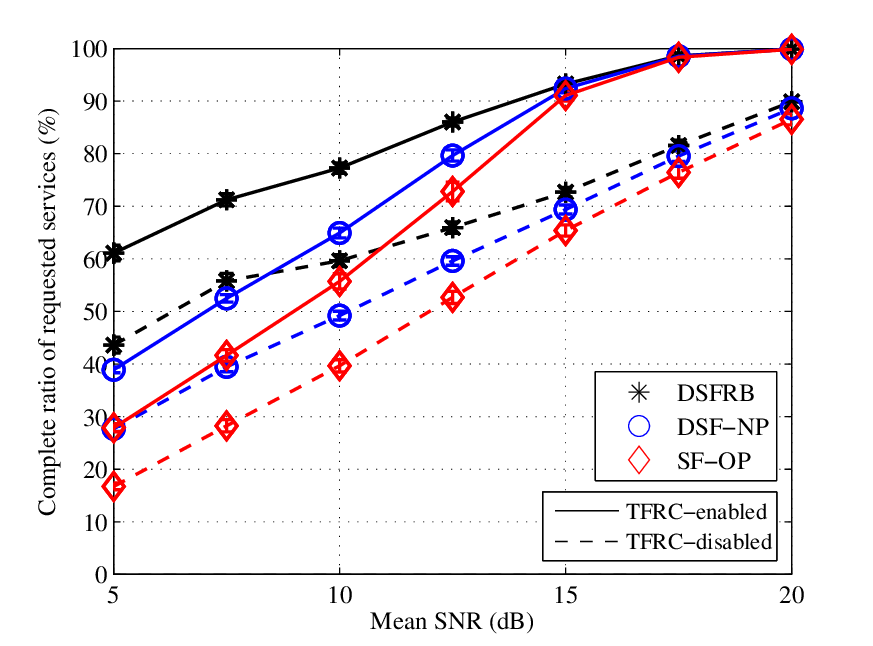}
\label{fig:snr offline:ratio}
}
\vspace{-0.3cm}
\caption{Impact of mean SNR on total system reward and complete ratio with offline scheduling.}
\vspace{-0.65cm}
\label{fig:snr offline}
\end{figure*}

In Fig. \ref{fig:lifetime offline}, we study the impact of mean request lifetime on the total system reward and the complete ratio of all requests, with ${\lambda}_{u}=0.075$ requests/s and $\mu \in \left[ 10,80 \right]$ s. %Other parameters are set the same as for Fig. \ref{fig:lambda offline}.
It is observed that both total system reward and complete ratio of the three algorithms, with or without TFRC, increase as the mean lifetime $\mu$ increases. %
With an increase of $\mu$, each request has more time for resource allocation, thus more chances to be delivered before deadline.
Yet, for both performance metrics, DSFRB is  much better than other two algorithms, due to the same reasons as aforesaid.
Also, by applying TFRC, all  algorithms perform better but tend to converge when $\mu \geq 40$ s. This is because the cell capacity has been fully exploited by the TFRC strategy when $\mu \geq 40$ s.
%with limited spectrum becomes saturated

%by applying TFRC, more accurate information of request deadline can be utilized in request scheduling, thus further improving the performance of the algorithms.

%From Fig. \ref{fig:lifetime offline} we can see that DSFRB algorithm has the best performance, DSF comes the second, and the worst is SF algorithm for old penalty.
%For instance, in Fig. \ref{fig:lifetime offline:ratio}, when implementing TFRC strategy, the performance gap between DSFRB (DSF-NP) and SF-OP is 20.91\% (8.29\%) respectively, showing the benefit of our newly proposed penalty method. Besides, DSFRB's performance improvement above DSF-NP also proves the efficiency of the algorithm's modification, which is mentioned as before.
%However, in both  Fig. \ref{fig:lifetime offline:reward} and  Fig. \ref{fig:lifetime offline:ratio}, the performance of DSFRB increases slowly and gradually comes to an upper bound, especially when $\mu \ge 40$. The reason is that the traffic load in this scene keeps the performance bound, and the increase of lifetime doesn't work due to the limited spectrum resource.

Fig. \ref{fig:snr offline} shows how the total system reward and the complete ratio of requested services change with the channel quality. Here, we set ${\lambda}_{u}=0.075$ requests/s, vary the mean SNR of each channel from 5 dB to 20 dB, and keep other parameters the same as for Fig. \ref{fig:lambda offline}. Both performance metrics improve with the channel quality because the data transmission rate increases with it as well. However, it is clear that integrating TFRC with scheduling techniques helps each algorithm harvest much more potential benefits from the improved channel quality.

%system reward and complete ratio of there algorithms increase with the increase of mean SNR, which represents the mean channel quality. The reason is that, with an increase of mean SNR, each subchannel transmits more data for the requests, thus much more requests can be transmitted before their deadlines. However, the performance gap among these three algorithms decreases as $\delta $ increases, especially when $\delta $ is large. Notice that, when $\delta =17.5$, the complete ratio of TFRC-enabled DSFRB is nearly 100\%, without much room for improvement. But other algorithms will have much performance improvement with a better channel condition. It also shows the advantage of TFRC strategy, by implementing which DSFRB provides an efficient scheduling solution.
\vspace{-0.25cm}
\begin{figure*}[htp]
\subfigcapskip = -7.5pt
\centering
\subfigure[Total reward vs. traffic load.]
{
\includegraphics[width=0.46\linewidth]{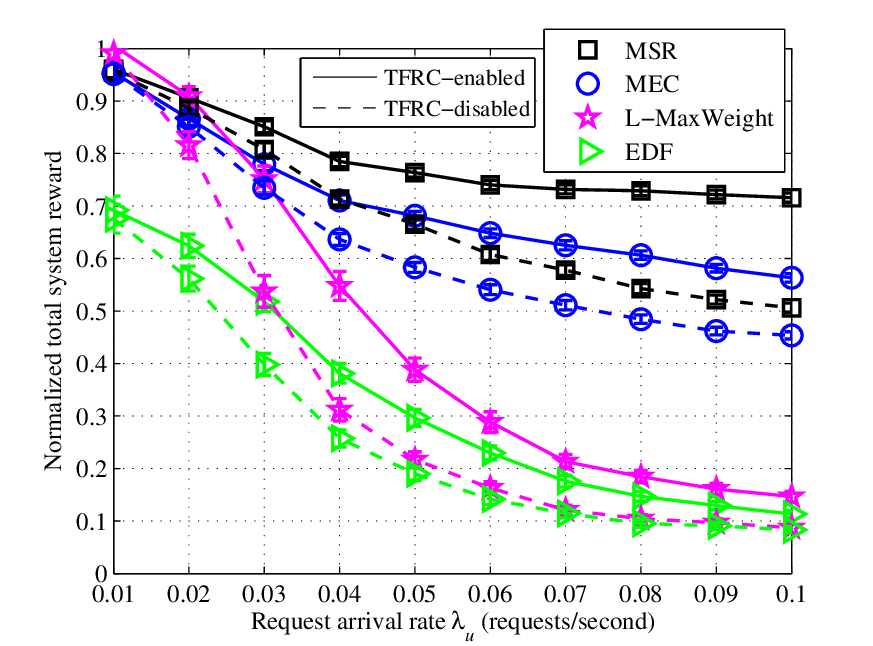}
\label{fig:lambda_online:reward}
}%
%\hspace{0.25in}
\subfigure[Complete ratio vs. traffic load.]
{
\includegraphics[width=0.46\linewidth]{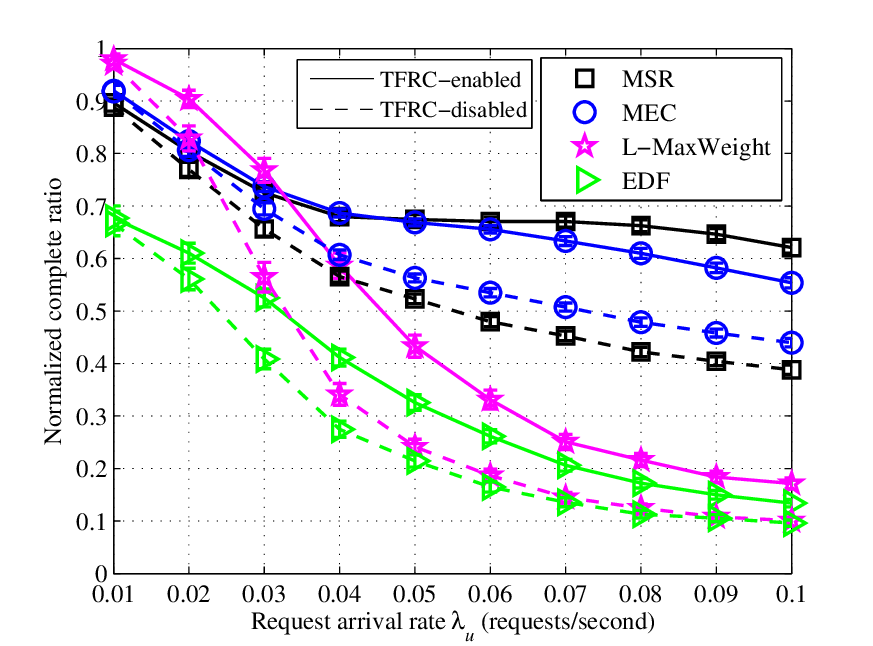}
\label{fig:lambda_online:ratio}
}
\vspace{-0.3cm}
\caption{Impact of traffic load on total system reward and complete ratio with online scheduling.}
\label{fig:lambda online}
\vspace{-0.5cm}
\end{figure*}

\subsection{Online Scheduling}

The good performance of TFRC-enabled schedule benefits from not only algorithm design but also non-causal information on request demand and channel capacity (see \cite{Shan_TWC_submitted}). Next, we evaluate its online counterpart, thus to understand the effect of TFRC-enabled scheduling techniques in a more practical scenario.
In the following, all simulation settings are the same as those for offline scheduling. The results are normalized by TFRC-enabled DSFRB algorithm.
%we give the performance curve of four online algorithms, including the proposed Smith ratio and exponential capacity algorithm, as well as EDF, L-Max Weight, a laxity-based opportunistic scheduling algorithm proposed in \cite{Wu_WCNC13}. All these performance curves are normalized by TFRC-enabled DSFRB algorithm. The parameters are the same as those in offline scheduling.

Fig. \ref{fig:lambda online} shows the impact of traffic load. % (request arrival rate ${{\lambda }_{\mu }}$) on both total system reward and complete ratio with online scheduling. The parameters are the same as those in Fig. \ref{fig:lambda offline}.
%Both the results for TFRC-enabled and TFRC-disabled online algorithms are presented.
%Again,
It is observed that each algorithm performs better if enabling TFRC. %, illustrating the benefit of TFRC in a practical scenario without full information.
As the traffic load increases, the normalized total system reward with any of the compared algorithms decreases, and the performance gap between these online algorithms and the DSFRB algorithm increases. %due to limited available information.
However, from the simulations we find that the total system reward (i.e., the absolute value) %for L-MaxWeight and EDF first increase when ${{\lambda }_{\mu }} \leq 0.03$ requests/s then decrease as traffic load keeps increasing, while those
with MSR or MEC always increases with the traffic load.
Further, the two algorithms perform much better than both L-MaxWeight and EDF, especially when traffic load is heavy. For example, when enabling TFRC and ${{\lambda }_{\mu }} = 0.1$ requests/s, as compared with L-MaxWeight the improvement of total system reward (complete ratio) of MSR and MEC can be 387.9\% and 283.8\% (262.1\% and 223.0\%), respectively.
However, TFRC-enabled L-MaxWeight performs slightly better than MSR and MEC when request arrival rate is less than 0.02 requests/s, showing the advantage of L-MaxWeight in a low traffic load region.

%Besides, it can be observed that there is a gap between the normalized performance of four online algorithms and the offline DSFRB algorithm.
%And this performance gap increases with traffic load.
%This is mainly due to the available information utilized in each scheduler, the offline DSFRB algorithm makes full use of the information of future requests and channel capacities, while the online EDF and L-MaxWeight algorithm only depends on request deadlines or the achievable multiuser diversity gain.
%Due to the predicted future channel information and other request information, such as request size, Smith ratio and exponential capacity algorithm's average performances outperform other two online algorithms in both total system reward and complete ratio.
%As shown in Fig. \ref{fig:lambda online}, the total system reward and complete ratio's performance gap between TFRC-enabled Smith ratio algorithm and EDF (L-MaxWeight) is 0.4598 (0.3309) and 0.3578 (0.2226) respectively.
%However, L-MaxWeight performs better than our proposed two algorithms when ${{\lambda }_{\mu }}\le 0.03$requests/s, the reason is that L-Max Weight is more applicable to the light traffic situation, but Smith ratio and exponential capacity algorithm are much better when the traffic load becomes heavier.

\vspace{-0.35cm}
\begin{figure*}[htp]
\subfigcapskip = -7.5pt
\centering
\subfigure[Total reward vs. average request size.]
{
\includegraphics[width=0.45\linewidth]{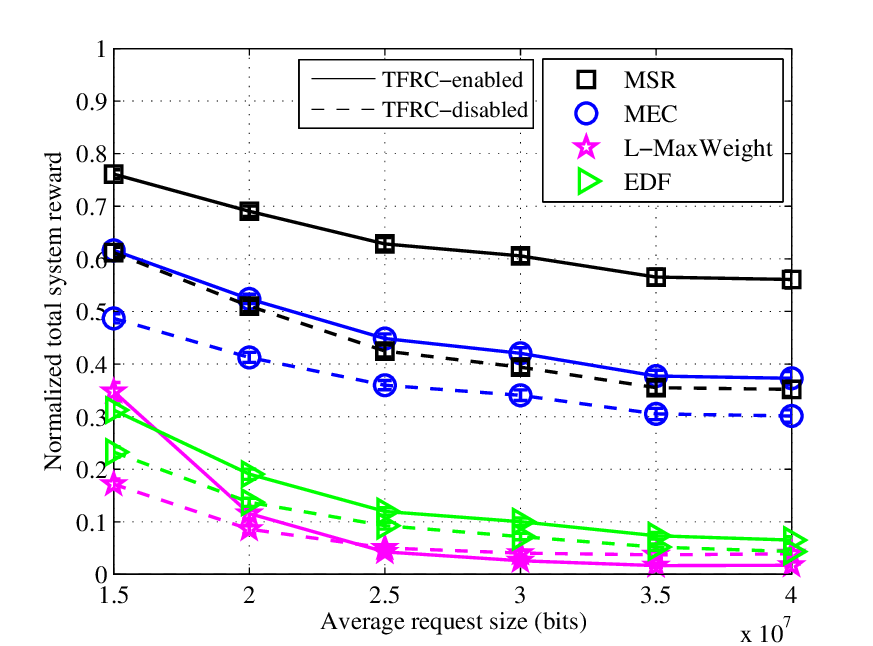}
\label{fig:data online:reward}
}%
%\hspace{0.25in}
\subfigure[Complete ratio vs. traffic load.]
{
\includegraphics[width=0.46\linewidth]{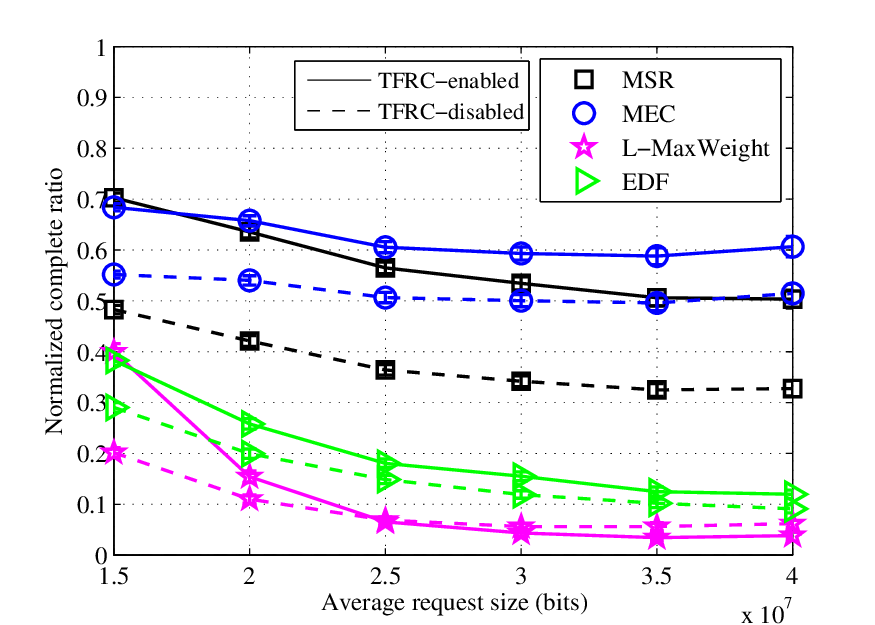}
\label{fig:data_online:ratio}
}
\vspace{-0.3cm}
\caption{Impact of average request size on total system reward and complete ratio with online scheduling.}
\label{fig:data online}
\vspace{-0.4cm}
\end{figure*}

Fig. \ref{fig:data online} evaluates the effect of request size.
The MSR and MEC outperform EDF and L-MaxWeight in both total system reward and complete ratio, as the average request size increases.
However, the normalized total system reward or complete ratio for each of the four algorithms reduces as the average request size increases, because of not only the increased traffic load but also the different scheduling capabilities of these algorithms as compared with the offline DSFRB.
Integrating with TFRC helps MSR, MEC, and EDF achieve much better performance than their TFRC-disabled counterparts, which does not hold for L-MaxWeight when average request size is larger than 2.5 Mbits.
The reason is that, with user behavior information, TFRC allows a request to be scheduled in a longer time, which also has an effect of accumulating network traffic load. The potentially increased traffic load generates a negative impact on L-MaxWeight.

%\vspace{-0.6cm}
\begin{figure*}[htp]
\subfigcapskip = -7.5pt
\centering
\subfigure[Total reward vs. mean request lifetime.]
{
\includegraphics[width=0.46\linewidth]{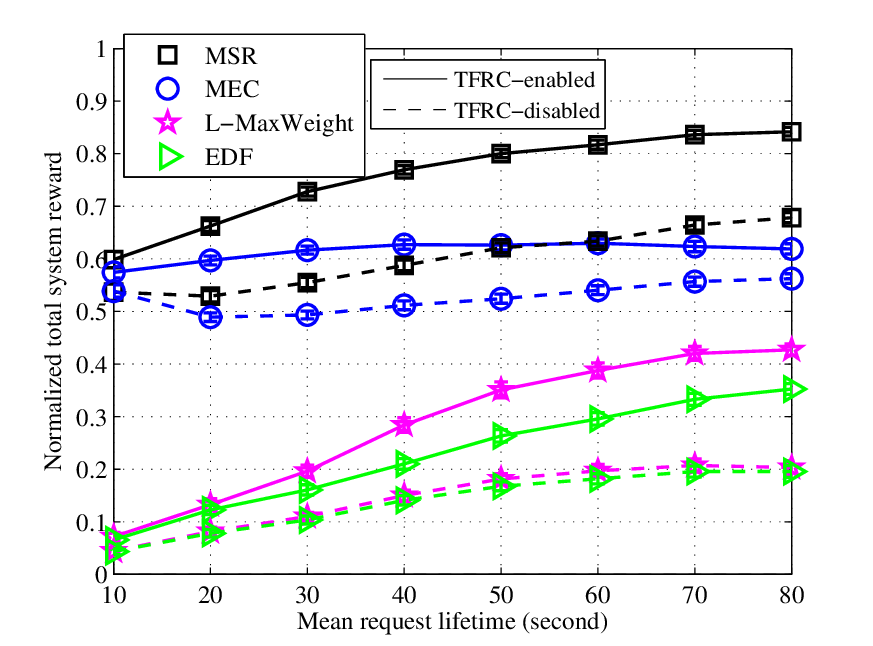}
\label{fig:lifetime_online:reward}
}%
%\hspace{0.25in}
\subfigure[Complete ratio vs. mean request lifetime.]
{
\includegraphics[width=0.46\linewidth]{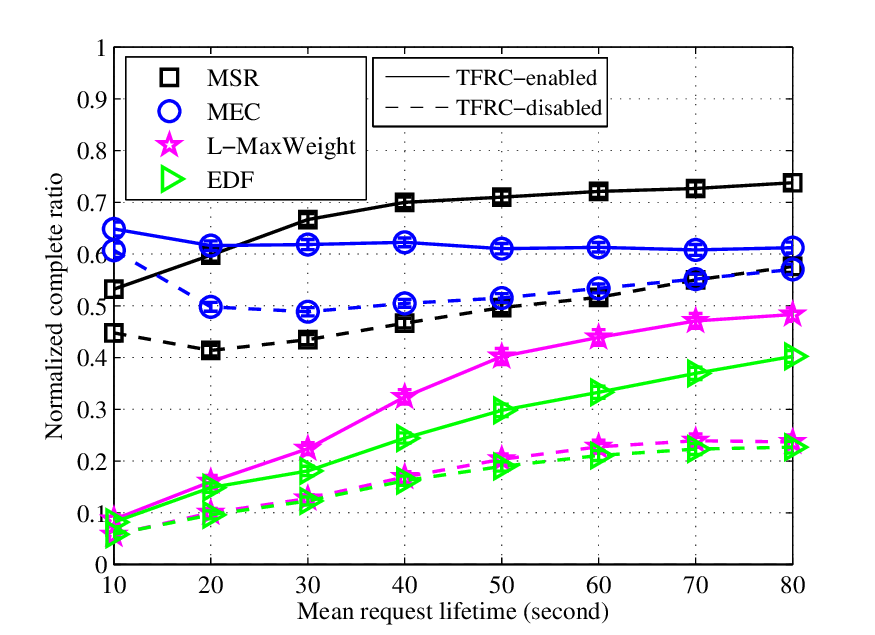}
\label{fig:lifetime_online:ratio}
}
\vspace{-0.3cm}
\caption{Impact of mean request lifetime on total system reward and complete ratio with online scheduling.}
\label{fig:lifetime_online}
\vspace{-0.55cm}
\end{figure*}

\begin{figure*}[tp]%Ë«À¼Í¼Æ¬¼Ó*
\subfigcapskip = -7.5pt
\centering
%\subfigbottomskip = 1pt
\subfigure[Total reward vs. mean SNR.]
{
\includegraphics[width=0.46\linewidth]{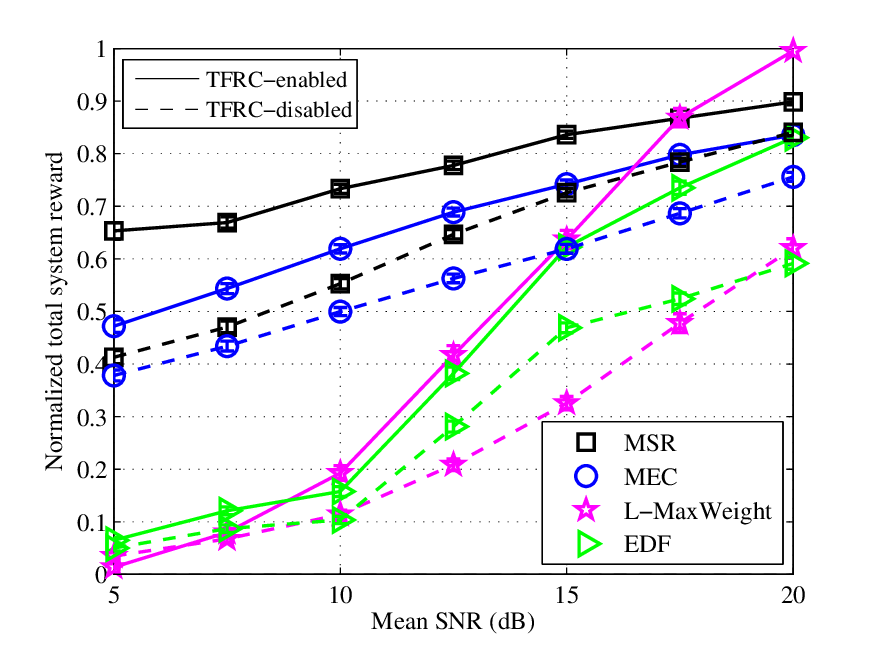}
\label{fig:snr online:reward}
}%
%\hspace{0.25in}
\subfigure[Complete ratio vs. mean SNR.]
{
\includegraphics[width=0.46\linewidth]{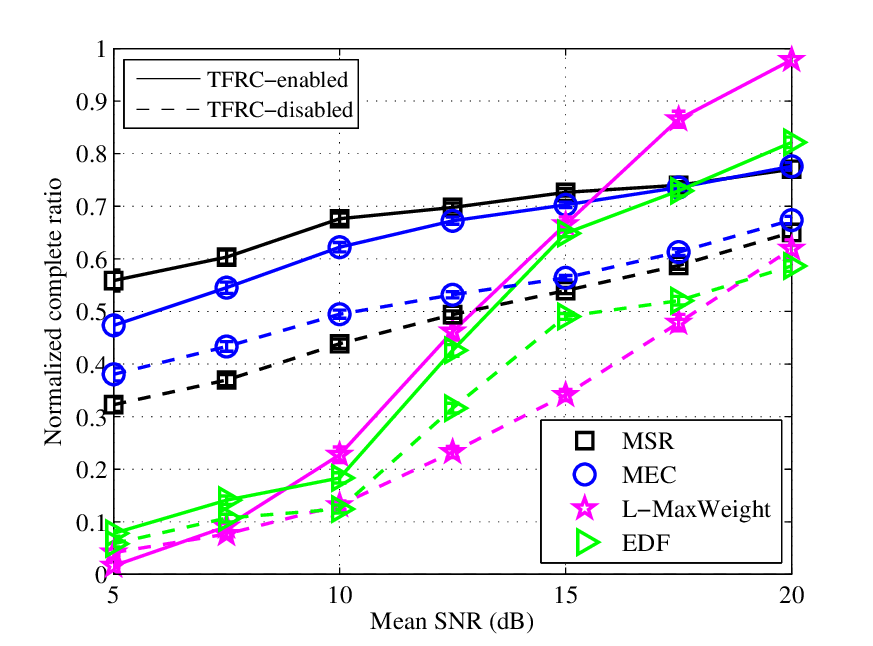}
\label{fig:snr online:ratio}
}
\vspace{-0.3cm}
\caption{Impact of mean SNR on total system reward and complete ratio with online scheduling.}
\vspace{-0.85cm}
\label{fig:snr online}
\end{figure*}

Fig. \ref{fig:lifetime_online} studies the impact of mean request lifetime. %on the total system reward and the complete ratio of all the requests for online scheduling with all the parameters are set the same as for Fig. \ref{fig:lifetime offline}.
It can be seen that the performance gap between the online algorithms and the offline DSFRB algorithm reduces with mean request lifetime. The performance loss in terms of the normalized total system reward is reduced by %from 0.46 to 0.32
30.4\%
for TFRC-disabled  MSR when mean request lifetime increases from 10 s to 80 s, implying the benefits of extending request scheduling time. %(further utilizing TFRC).
Besides, this improvement is more obvious if TFRC is enabled, e.g., the same performance loss can be further reduced %from 0.4 to 0.16
by 60\%
for TFRC-enabled MSR when mean request life time changes within the same range.
In Fig. \ref{fig:lifetime_online}, although MEC behaves different from other algorithms in terms of either normalized performance metric,
the absolute values of both performance metrics with any of the compared algorithms increase as mean request lifetime extends.
The dissimilarity is due to the different performance improvement speeds with respect to mean request time among those online and the offline DSFRB algorithms.

Fig. \ref{fig:snr online} shows the impact of channel quality on both total system reward and complete ratio with online scheduling.
The normalized reward and complete ratio increase as the mean SNR increases.
This is because the improved channel quality greatly shortens the time to finish a request, which compensates for the negative impact of lacking accurate information on request demand and channel capacity, as compared with the offline DSFRB algorithm.
Yet, as shown in the figure, when the mean SNR is larger than 17.5 dB, TFRC-enabled L-MaxWeight performs much better than the two newly proposed online algorithms. %MSR and MEC.
This is rational since the wireless network with the good channel quality can accommodate a traffic load larger than the one set in the simulations, i.e., corresponding to an underloaded system.
Nevertheless, the proposed two algorithms perform much better when channel quality is poor or network traffic load is heavy.
%%

%=======================================end comments==============================================


\begin{thebibliography}{10}
%\fontsize{7.45}{7.45}\selectfont

\providecommand{\url}[1]{#1} \csname url@rmstyle\endcsname
\providecommand{\newblock}{\relax} \providecommand{\bibinfo}[2]{#2}
\providecommand\BIBentrySTDinterwordspacing{\spaceskip=0pt\relax}
\providecommand\BIBentryALTinterwordstretchfactor{4}
\providecommand\BIBentryALTinterwordspacing{\spaceskip=\fontdimen2\font
plus \BIBentryALTinterwordstretchfactor\fontdimen3\font minus
  \fontdimen4\font\relax}
\providecommand\BIBforeignlanguage[2]{{%
\expandafter\ifx\csname l@#1\endcsname\relax
\typeout{** WARNING: IEEEtran.bst: No hyphenation pattern has been}%
\typeout{** loaded for the language `#1'. Using the pattern for}%
\typeout{** the default language instead.}%
\else \language=\csname l@#1\endcsname \fi #2}}

\bibitem{Shan_TWC_submitted}
H.~Shan, Y.~Zhang, W.~Zhuang, A.~Huang, and Z.~Zhang, ``User behavior-aware scheduling based on time-frequency resource conversion," submitted to \emph{ IEEE Trans. Vehicular Technology}.

\bibitem{1zhang}
Y.~Zhang, H.~Shan, W.~Zhuang, and A.~Huang, ``Time-frequency resource conversion based scheeduling for on-demand data services," in \emph{Proc. IEEE Globecom}, 2015.

\bibitem{Wu_WCNC13}
H.~Wu, X.~Liu, and Y.~Zhang, ``Laxity-based opportunistic scheduling with flow-level dynamics and deadlines," in \emph{Proc. IEEE WCNC}, 2013.


\end{thebibliography}
\end{document}